\documentclass[a4paper,12pt]{article}

\pdfoutput=1
\usepackage{amsmath}
\usepackage{amssymb}
\usepackage{amsfonts}
\usepackage{mathrsfs}
\usepackage{bbm}
\usepackage{graphicx,subfigure}

\usepackage{bookmark}

\usepackage{cite}

\usepackage{calc}
\usepackage{hyperref}
\usepackage{array}

\usepackage{ulem}

\usepackage{multirow}
\usepackage{color}
\usepackage{xcolor,colortbl}

\usepackage{psfrag}
\usepackage{pstricks}
\usepackage{epsfig}

\allowdisplaybreaks

\newlength{\dinwidth}
\newlength{\dinmargin}
\definecolor{nicered}{rgb}{1.0,0.0,0.2}

\definecolor{color1}{rgb}{0.9,.4,.2}

\definecolor{color2}{rgb}{0.3,.6,.7}

\definecolor{color3}{rgb}{0.7,.2,.7}

\usepackage{color}

\begin{document}

\title{
\vspace*{-0.5cm}
\bf \Large
Study of  the isospin breaking decay $\mathbf{\boldsymbol{Y(2175)\rightarrow\phi f_0(980)\rightarrow\phi\eta{\pi}^0}}$ at BESIII}

\author{Xiao-Dong Cheng$^{1}$\footnote{chengxd@mails.ccnu.edu.cn}, Hai-Bo Li$^{2,3}$\footnote{lihb@ihep.ac.cn}, Ru-Min Wang$^{1}$\footnote{ruminwang@sina.com}, Mao-Zhi Yang$^{4}$\footnote{yangmz@nankai.edu.cn}\\
\\
{$^1$\small College of Physics and Electronic Engineering,}\\[-0.2cm]
{    \small Xinyang Normal University, Xinyang 464000, People's Republic of China}\\[-0.1cm]
{$^2$\small  Institute of High Energy Physics,}\\[-0.2cm]
{    \small  Beijing 100049, People's Republic of China}\\[-0.1cm]
{$^3$\small  University of Chinese Academy of Sciences,}\\[-0.2cm]
{    \small  Beijing 100049, People's Republic of China}\\[-0.1cm]
{$^4$\small School of Physics,}\\[-0.2cm]
{    \small Nankai University, Tianjin 300071, People's Republic of China}\\[-0.1cm]}

\date{}
\maketitle
\bigskip\bigskip
\maketitle
\vspace{-1.2cm}

\begin{abstract}
Using  measured branching fraction of the decay $J/\psi \rightarrow\eta Y(2175))\rightarrow\eta\phi f_0(980)\rightarrow\eta\phi\pi^+\pi^{-}$ from the BESIII experiment, we estimate branching fraction of $J/\psi \rightarrow\eta Y(2175))\rightarrow\eta\phi f_0(980)\rightarrow\eta\phi\eta\pi^{0}$ decay,  which proceeds via the $f_0(980)$-$a_0^0(980)$ mixing and the $\pi^0$-$\eta$ mixing. The branching fraction is predicted to be about $O(10^{-6})$, which can be accessed with $10^{10}$ $J/\psi$ events collected at the BESIII. The decay is dominated by the contribution from $f_0(980)$-$a_0^0(980)$ mixing. We find that the interference between the amplitudes due to  $f_0(980)$-$a_0^0(980)$ mixing and that due to $\pi^0$-$\eta$ mixing is destructive. The branching fraction can be decreased by about $10\%$ owing to the interference effect.
We also study the $\eta\pi^0$ mass squared spectrum,  and find that a narrow peak due to the $f_0(980)$-$a_0^0(980)$ mixing in the $\eta\pi^0$ mass squared spectrum should  be observed.
The observation of this decay in experiment will be  helpful to determine the $f_0(980)$-$a_0^0(980)$ mixing intensity and get information about the structures of the light scalar mesons.

\end{abstract}
\newpage

\section{Introduction}
\label{sec:intro}
The nature of the light scalar mesons $a_0^0(980)$ and $f_0(980)$ is still a hot topic in hadronic physics. Several models about the structure of the scalar mesons have been proposed, such as $q \bar{q}$ states, glueball, hybrid states, molecule states, tetra-quark states and the superpositions of these contents~\cite{Cheng:2005nb, Weinstein:1982gc, Weinstein:1983gd, Weinstein:1990gu, Jaffe,Kim:2017yur,Abdel-Rehim:2014zwa, Amsler:1995td, Amsler:1995tu, Amsler:2002ey,Gorishnii:1983zi}. Due to the absence of convincing evidence, a final consensus has not been reached so far. Therefore, more researches both in theory and experiment are still needed.

The structure of $a_0^0(980)$ and $f_0(980)$ is closely related to the mixing of them, which was first suggested theoretically in Ref.~\cite{Achasov:1979xc}. Its mixing intensity has been studied extensively on its different aspects and possible manifestations in various processes~\cite{Kudryavtsev:2001ee,Achasov:2004ur,Achasov:2003se,Wu:2007jh,Hanhart:2007bd,Wu:2008hx,Aceti:2012dj,Roca:2012cv,Aceti:2015zva,Achasov:2015uua,Wang:2016wpc,Achasov:2016wll,Achasov:2017edm,Achasov:2017zhu,Sakai:2017iqs,Bayar:2017pzq,Achasov:2017ncx,Liang:2017ijf}. Recently, BESIII Collaboration has reported the first observation of  $f_0(980)-a_0^0(980)$ mixing in the decays of $J/\psi \rightarrow \phi f_0(980)\rightarrow \phi a_0^0(980)\rightarrow\phi\eta\pi^0$ and $\chi_{c1}\to a^{0}_{0}(980)\pi^{0}\to f_{0}(980)\pi^{0}\to\pi^{+}\pi^{-}\pi^{0}$  \cite{Ablikim:2018pik}. In their work, the values of the mixing intensity $\xi_{fa}$  for the $f_0(980)-a_0^0(980)$ transition was obtained
\begin{align}\label{eq:epsilonfabesiii}
  \begin{split}
     \xi_{fa} &= (0.99 \pm 0.35) \times 10^{-2} ~~~ \text{(solution-1)}~, \\
     \xi_{fa} &= (0.41 \pm 0.25) \times 10^{-2} ~~~ \text{(solution-2)}~.
   \end{split}
\end{align}
Here, $\xi_{fa}$ is defined as
\begin{align}\label{eq:epsilonfadefine}
\xi_{fa}=\frac{{\mathcal B}(J/\psi \rightarrow \phi f_0(980)\rightarrow \phi a_0^0(980)\rightarrow \phi \eta \pi^0)}{{\mathcal B}(J/\psi \rightarrow \phi f_0(980)\rightarrow \phi \pi^{+} \pi^{-})}.
\end{align}
The theoretical calculation prefers to the solution-1 result of BESIII~\cite{Aliev:2018bln}. Here, more works are needed to determine the final solution of $\xi_{fa}$.

The $Y(2175)$ resonance, which deays dominantly via a $\phi f_0(980)$ intermediate state, is a vector meson, its ${J^{PC}}=1^{--}$~\cite{Shen:2009mr}. This resonance was first observed by BABAR Collaboration~\cite{Aubert:2006bu} and then confirmed by BESIII Collaboration~\cite{Ablikim:2007ab} and Belle Collaboration~\cite{Shen:2009zze}. Recent result on $Y(2175)$ resonance in $J/\psi$ decay from BESIII Collaboration is obtained as~\cite{Ablikim:2014pfc}
\begin{align}\label{eq:besiiijpsiy2175}
{\mathcal B}(J/\psi \rightarrow\eta Y(2175)\rightarrow\eta \phi f_0(980)\rightarrow \eta \phi\pi^{+} \pi^{-})
 =(1.20\pm0.40)\times 10^{-4}.
\end{align}

In this paper, we study the isospin breaking decay ${J/\psi \rightarrow\eta Y(2175)\rightarrow\eta \phi f_0(980)\rightarrow\eta \phi\eta{\pi}^0}$ and estimate its branching fraction by using recent measurements by the BESIII~\cite{Ablikim:2018pik,Ablikim:2014pfc}. We also study the distribution of  $\eta\pi^0$ mass squared spectrum near the $K\bar{K}$ threshold.

\section{Two mechanisms of the decay}
\label{sec:Two mechanisms of the decay}
\begin{figure}[t]
\centering
\includegraphics[width=0.6\textwidth]{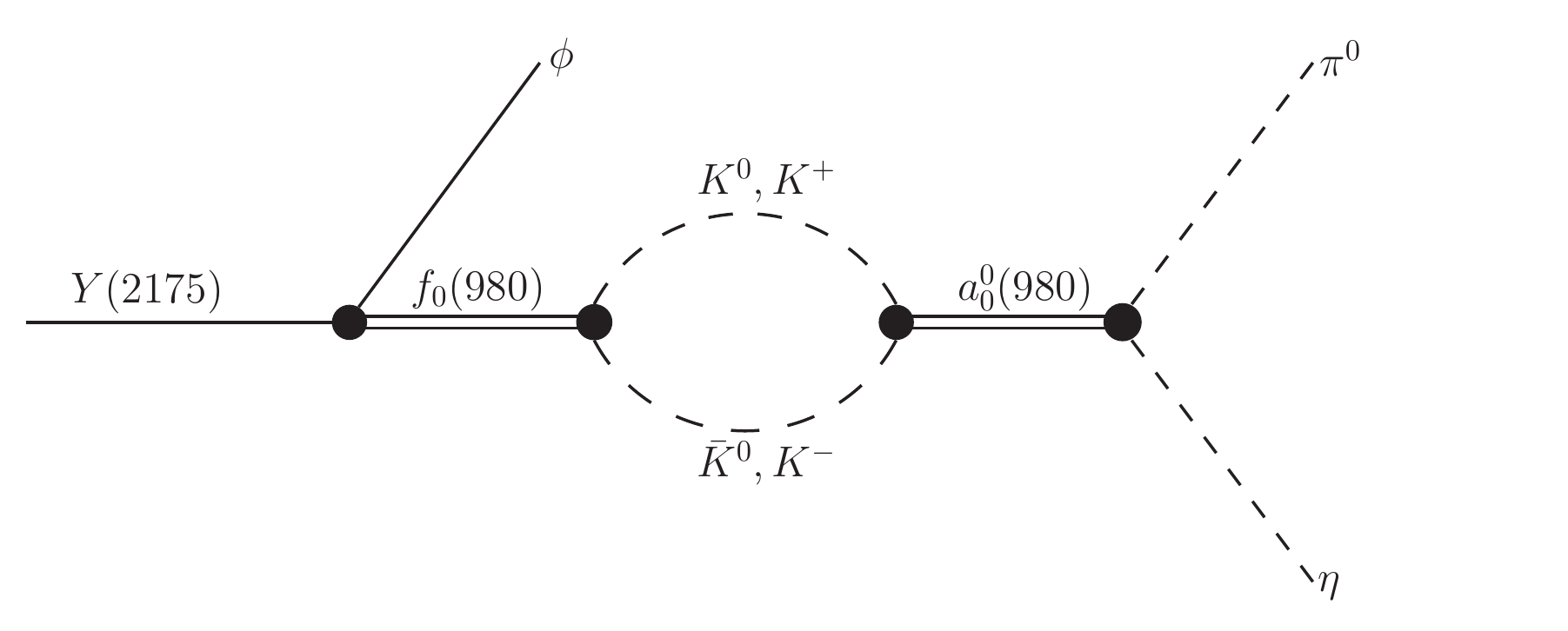}
\caption{\small Feynman diagram for the reaction $Y(2175)\rightarrow\phi f_0(980)\rightarrow\phi a_0^0(980)\rightarrow\phi\eta\pi^{0}$.}
\label{yphifamixing}
\end{figure}

\begin{figure}[t]
\centering
\includegraphics[width=0.42\textwidth]{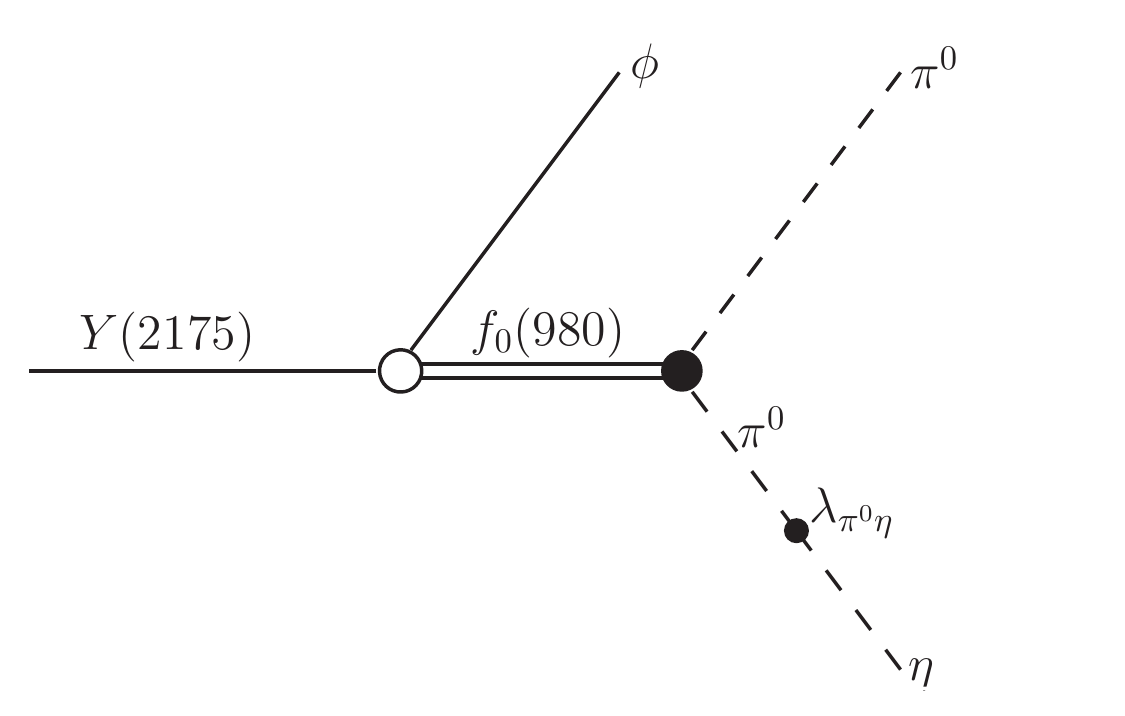}
\caption{\small Feynman diagram for the reaction $Y(2175)\rightarrow\phi f_0(980)\rightarrow\phi  \pi^0\pi^0\rightarrow\phi\eta\pi^{0}$.}
\label{yphietapimixing}
\end{figure}

The isospin breaking decay ${J/\psi \rightarrow\eta Y(2175)\rightarrow\eta \phi f_0(980)\rightarrow\eta \phi\eta{\pi}^0}$ can proceed via the $f_0(980)$-$a_0^0(980)$ mixing and the $\pi^0$-$\eta$ mixing. The amplitude can be written as
\begin{align}\label{eq:famixingandpietamixing}
&{\mathcal M}(J/\psi \rightarrow\eta Y(2175)\rightarrow\eta\phi f_0(980)\rightarrow \eta\phi\eta\pi^{0})=\nonumber
\\&~~~~={\mathcal M}(J/\psi \rightarrow\eta Y(2175)\rightarrow\eta\phi f_0(980)\rightarrow\eta\phi a_0^0(980)\rightarrow \eta\phi\eta\pi^{0})\nonumber
\\&~~~~~~~+{\mathcal M}(J/\psi \rightarrow\eta Y(2175)\rightarrow\eta\phi f_0(980)\rightarrow\eta\phi \pi^0\pi^0\rightarrow \eta\phi\eta\pi^{0}),
\end{align}
 The corresponding graphs are shown in Figs.~\ref{yphifamixing} and \ref{yphietapimixing}. For the contribution of $f_0(980)$-$a_0^0(980)$ mixing, the mixing intensity $\xi_{fa}$ can be expressed in a similar way as that in Eq.(\ref{eq:epsilonfadefine}) , and here is defined as
\begin{align}\label{eq:epsilfadefiney2175}
\xi_{fa}=\frac{{\mathcal B}(J/\psi \rightarrow\eta Y(2175)\rightarrow\eta\phi f_0(980)\rightarrow\eta\phi a_0^0(980)\rightarrow \eta\phi\eta\pi^{0})}{{
\mathcal B}(J/\psi \rightarrow\eta Y(2175)\rightarrow\eta\phi f_0(980)\rightarrow\eta\phi \pi^+\pi^-)}.
\end{align}
Combining Eqs.(\ref{eq:epsilonfabesiii}), (\ref{eq:besiiijpsiy2175}) and (\ref{eq:epsilfadefiney2175}),
one can obtain the branching fraction of $J/\psi \rightarrow\eta Y(2175)\rightarrow\eta\phi f_0(980)
\rightarrow\eta\phi a_0^0(980)\rightarrow \eta\phi\eta\pi^{0}$.

With $\xi_{fa}=(0.99\pm 0.35)\times 10^{-2}$, one can obtain
\begin{equation}
{\mathcal B}(J/\psi \rightarrow\eta Y(2175)\rightarrow\eta\phi f_0(980)\rightarrow\eta\phi a_0^0(980)\rightarrow \eta\phi\eta\pi^{0})=(1.19\pm 0.58)\times 10^{-6}\label{eq:branchratioYphifaetapi1},
\end{equation}
while, with $\xi_{fa}=(0.41\pm 0.25)\times 10^{-2}$, one can get
\begin{equation}
{\mathcal B}(J/\psi \rightarrow\eta Y(2175)\rightarrow\eta\phi f_0(980)\rightarrow\eta\phi a_0^0(980)\rightarrow \eta\phi\eta\pi^{0})=(0.49\pm 0.34)\times 10^{-6}\label{eq:branchratioYphifaetapi2}.
\end{equation}
In BESIII analysis for the decays of $J/\psi \rightarrow \phi f_0(980)\rightarrow \phi a_0^0(980)\rightarrow\phi\eta\pi^0$ and $\chi_{c1}\to a^{0}_{0}(980)\pi^{0}\to f_{0}(980)\pi^{0}\to\pi^{+}\pi^{-}\pi^{0}$  \cite{Ablikim:2018pik},  they only assumed contribution from $f_0(980)-a_0^0(980)$ or $a_0^0(980)-f_0(980)$ mixing, which causes the  isospin breaking decays.   In fact,  the final states of $\phi \eta \pi^0$ could be also induced by
$J/\psi \rightarrow \phi f_0(980)\rightarrow \phi \pi^0\pi^0 \rightarrow\phi\eta\pi^0$ via  $\pi^0$-$\eta$ mixing.  If it is the case that the isospin breaking decay is due to both $f_0(980)-a_0^0(980)$ and  $\pi^0$-$\eta$ mixings,   actually the BESIII measured values for the mixing intensity $\xi_{fa}$ in Eq. (\ref{eq:epsilonfabesiii}) has already included both effects.
Therefore the results given in Eqs. (\ref{eq:branchratioYphifaetapi1}) and (\ref{eq:branchratioYphifaetapi2}) includes both the contributions of $f_0(980)$-$a_0^0(980)$ mixing and $\pi^0$-$\eta$ mixing physically.

As for the sole contribution of $\pi^0$-$\eta$ mixing, the relative  ratio of  ${\mathcal B}(f_0(980)\rightarrow\pi^{0}\pi^{0}\rightarrow\eta\pi^{0}$) to ${\mathcal B}(f_0(980)\rightarrow\pi^{0}\pi^{0}$) is
\begin{align}\label{eq:ratiofetapipipi1}
&\frac{{\mathcal B}(f_0(980)\rightarrow\pi^{0}\pi^{0}\rightarrow\eta\pi^{0})}{{\mathcal B}( f_0(980)\rightarrow\pi^{0}\pi^{0})}=4\frac{f(m_{f_0},m_{\eta},m_{\pi^0})} {f(m_{f_0},m_{\pi^0},m_{\pi^0})}\left|{\frac{\lambda_{\pi^0\eta}}{m_{\eta}^2-m_{\pi^0}^2}}\right|^2.
\end{align}
where $m_{f_0},m_{\eta},m_{\pi^0}$ are the masses of $f_0(980)$, $\eta$ and $\pi^0$ , respectively. The function $f$ is
\begin{equation}
f(x,y,z)=\sqrt{{x^4} +{y^4}+{z^4}-2{x^2}{y^2}-2{x^2}{z^2}- 2{y^2}{z^2} }.
\end{equation}
$\lambda_{\pi^0\eta}$ is the $\pi^0$-$\eta$ transition amplitude~\cite{Kudryavtsev:2001ee,Niskanen:1998yi}, which can be extracted from the  ratio of ${\mathcal B}(\eta^{\prime}\rightarrow\pi^+\pi^-\pi^0)$ and ${\mathcal B}(\eta^{\prime}\rightarrow\pi^+\pi^-\eta)$ decays~\cite{Coon:1986ie}
\begin{align}\label{eq:ratiofetapipipi2}
&\left|{\frac{\lambda_{\pi^0\eta}}{m_{\eta}^2-m_{\pi^0}^2}}\right|^2=\frac{{\mathcal B}(\eta^{\prime}\rightarrow \pi^+\pi^-\pi^0)}{{\mathcal B}(\eta^{\prime}\rightarrow\pi^+\pi^-\eta)}\frac{\phi_s(\eta^{\prime}\rightarrow\pi^+\pi^-\eta)}{\phi_s(\eta^{\prime}\rightarrow\pi^+\pi^-\pi^0)},
\end{align}
where  $\phi_s(\eta^{\prime}\rightarrow\pi^+\pi^-\eta)=\int_{4 m_{\pi^+}^2}^{(m_{\eta^{\prime}}-m_{\eta})^2} \frac{dq^2}{q^2} f(m_{\eta^{\prime}},\sqrt{q^2},m_{\eta})f(\sqrt{q^2},m_{\pi^+},m_{\pi^+})$, is the phase-space integral. While $\phi_s(\eta^{\prime}\rightarrow\pi^+\pi^-\pi^0)$ is the relevant phase-space integral that changes $m_{\eta}$ to $m_{\pi^0}$ in $\phi_s(\eta^{\prime}\rightarrow\pi^+\pi^-\eta)$. The relative ratio of ${\mathcal B}(\eta^{\prime}\rightarrow \pi^+\pi^-\pi^0)/{\mathcal B}(\eta^{\prime}\rightarrow\pi^+\pi^-\eta)$ has been measured by BESIII ~\cite{BESIII:2012aa} and CLEO Collaboration~\cite{Naik:2008aa}. The recent value measured by BESIII Collaboration is $(8.8\pm1.2)\times 10^{-3}$~\cite{Ablikim:2016frj,Fang:2017qgz}. Employing the relation ${\mathcal B}(f_0(980)\rightarrow\pi^{+}\pi^{-})=2{\mathcal B}(f_0(980)\rightarrow\pi^{0}\pi^{0})$, combining Eqs.(\ref{eq:besiiijpsiy2175}), (\ref{eq:ratiofetapipipi1}), (\ref{eq:ratiofetapipipi2}) and using the particle masses taken from Table~\ref{massofpietak} and Table~\ref{inputmassparameter}, one can obtain
\begin{align}\label{eq:branchratioYphifetapimxing}
{\mathcal B}(J/\psi \rightarrow\eta Y(2175)\rightarrow\eta\phi f_0(980)\rightarrow\eta\phi \pi^0\pi^0\rightarrow \eta\phi\eta\pi^{0})=(0.86\pm 0.31)\times 10^{-7}.
\end{align}
Obviously, this is much smaller than the results given in Eqs. (\ref{eq:branchratioYphifaetapi1}) and (\ref{eq:branchratioYphifaetapi2}), which implies that the contribution of $f_0(980)$-$a_0^0(980)$ mixing dominates over that of $\pi^0$-$\eta$ mixing in decay ${J/\psi \rightarrow\eta Y(2175)\rightarrow\eta \phi f_0(980)\rightarrow\eta \phi\eta{\pi}^0}$.

\begin{table}[t]
\begin{center}
\caption{\label{massofpietak} \small The masses of the particles in final states. }
\vspace{0.2cm}
\doublerulesep 0.8pt \tabcolsep 0.18in
\begin{tabular}{cc}
\hline
$m_{\pi^+}=  139.6 MeV$ ~\cite{Tanabashi:2018oca}                         &$m_{\pi^0}=  135 MeV$ ~\cite{Tanabashi:2018oca}  \\
\hline
$m_{K^+}=  493.7 MeV$  ~\cite{Tanabashi:2018oca}                        &$m_{K^0}=  497.6 MeV$ ~\cite{Tanabashi:2018oca}     \\
\hline
$m_{\eta}=  547.9 MeV$~\cite{Tanabashi:2018oca}                &$m_{\eta^{\prime}}=  (957.8\pm 0.1) MeV$~\cite{Tanabashi:2018oca}  \\
\hline
\end{tabular}
\end{center}
\end{table}

\section{The branching fraction }
\label{sec:The branching ratio}
\begin{figure}[t]
\centering
\includegraphics[width=0.42\textwidth]{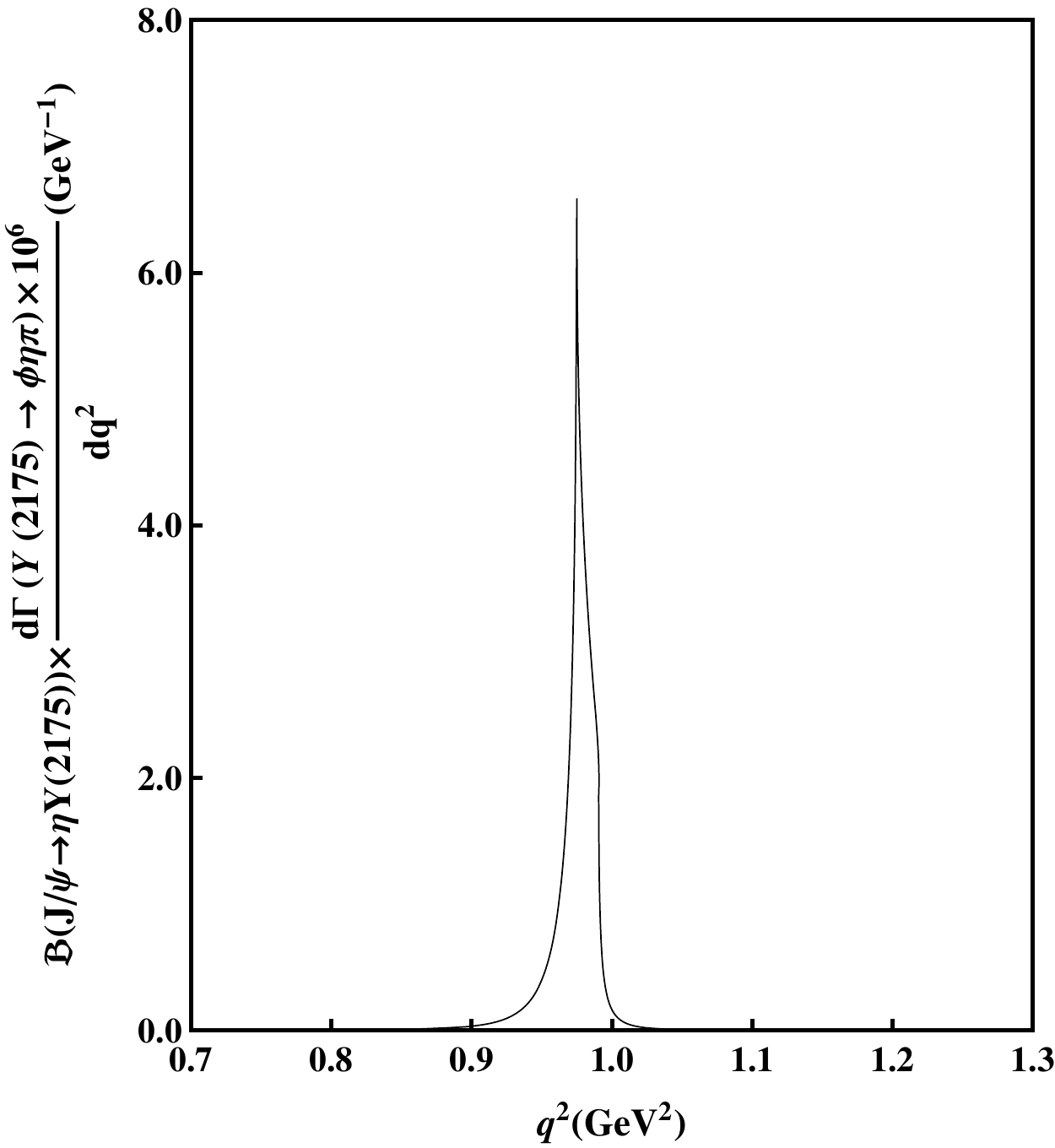}
\caption{\small the distribution of the $\eta\pi^0$ mass squared spectrum  $(q^2=(p_\eta+p_{\pi^0})^2)$ for the decay $J/\psi \rightarrow\eta Y(2175))\rightarrow\eta\phi f_0(980)\rightarrow\eta\phi \eta\pi^{0}$.}
\label{maffsquaredistribution}
\end{figure}

As mentioned in Eq.(\ref{eq:famixingandpietamixing}), both the $f_0(980)$-$a_0^0(980)$ mixing and the $\pi^0$-$\eta$ mixing can contribute to the decay of $J/\psi \rightarrow\eta Y(2175)\rightarrow\eta\phi f_0(980)\rightarrow \eta\phi\eta\pi^{0}$. The most characteristic feature of the first contribution is the narrow peak in the $\eta\pi^0$ mass spectrum, which is due to the property of the $f_0(980)$-$a_0^0(980)$ mixing amplitude~\cite{Achasov:1979xc,Wu:2007jh,Achasov:2017ncx}.  As far as $\pi^0$-$\eta$ mixing is concerned, however, the width in the $\eta\pi^0$ mass spectrum should  be  the natural width of $f_0(980)$ state, which is broad. Fortunately, the contribution from the $\pi^0$-$\eta$ mixing is much smaller than that from the $f_0(980)$-$a_0^0(980)$ mixing, so the narrow structure caused by the $f_0(980)$-$a_0^0(980)$ mixing is expected to be observed, while the broad width from the effect of $\pi^0$-$\eta$ mixing is negligibly small.  The corresponding decay amplitude contributed by  $f_0(980)$-$a_0^0(980)$ mixing is~\cite{Achasov:2017ncx,Achasov:2018grq}
\begin{align}\label{eq:decayamplitudeoffamixng}
&{\mathcal M}(Y(2175)\rightarrow\phi f_0(980)\rightarrow\phi a_0^0(980)\rightarrow\phi\eta\pi^{0})=\nonumber\\
&~~={\mathcal M}(Y(2175)\to\phi f_0(980))\cdot\frac{\Pi_{a_0 f_0}(q^2)}{D_{a_0}(q^2)D_{f_0}(q^2)-\Pi_{a_0 f_0}^2(q^2)}
\cdot g_{a_0\eta\pi^0},
\end{align}
where $q^2=(p_\eta+p_{\pi^0})^2$, and $g_{a_0\eta\pi^0}$ is the coupling of $a_0^0(980)$ to $\eta\pi^0$. ${\mathcal M}(Y(2175)\to \phi f_0(980))$ is the invariant amplitude for the decay $Y(2175)\rightarrow\phi f_0(980)$, which can be used to calculate the branching fraction
\begin{align}\label{eq:decayamplitudeofyfphi}
&{\mathcal B}(Y(2175)\rightarrow\phi f_0(980))=
\left|{\mathcal M}(Y(2175)\to\phi f_0(980))\right|^2\cdot\frac{f(m_{Y},m_{\phi},m_{f_0})}{16\pi\Gamma_{Y}m_{Y}^3},
\end{align}
where $\Gamma_{Y}$ is the decay width of $Y(2175)$. $m_Y$, $m_\phi$ and $m_{f_0}$ are the masses of the resonances $Y(2175)$, $\phi$ and $f_0(980)$, respectively. $\Pi_{a_0 f_0}(q^2)$ is the $f_0(980)$-$a_0^0(980)$ mixing amplitude, and defined as
\begin{align}\label{eq:decaywidthoffamixng2}
\Pi_{a_0 f_0}(q^2)=&\frac{g_{a_0 K^{+}K^{-}}g_{f_0 K^{+}K^{-}}}{16\pi}\bigg[i\left(R_{K^{+}K^{-}}(q^2)-R_{K^{0}{\bar{K}}^{0}}(q^2)\right)\bigg.\nonumber\\
&\bigg.-\frac{R_{K^{+}K^{-}}(q^2)}{\pi} \ln{\frac{1+R_{K^{+}K^{-}}(q^2)}{1-R_{K^{+}K^{-}}(q^2)}}+\frac{R_{K^{0}{\bar{K}}^{0}}(q^2)}{\pi} \ln{\frac{1+R_{K^{0}{\bar{K}}^{0}}(q^2)}{1-R_{K^{0}{\bar{K}}^{0}}(q^2)}}\bigg],
\end{align}
where for $q^2>4 m_a^2$, $R_{aa}(q^2)=\sqrt{1-4m_a^2/q^2}$, while for $0<q^2\leq4 m_a^2$, $R_{aa}(q^2)=i\sqrt{4m_a^2/q^2-1}$, here $a=K^\pm,K^0$. $D_r(q^2)$ in Eq.(\ref{eq:decayamplitudeoffamixng}) is the denominator for the propagator of the resonance $r$,
\begin{align}\label{eq:decaywidthoffamixng3}
D_{r}(q^2)=q^2-m_r^2-\sum\limits_{ab}\left[\text{Re}\Pi_r^{ab}(m_r^2)-\Pi_r^{ab}(q^2)\right].
\end{align}
For $r=a_0^0(980)$, $ab=\left(\eta\pi^0,K^+K^-,K^0{\bar{K}}^0\right)$, and for $r=f_0(980)$, $ab=\left(\pi^+\pi^-,\pi^0\pi^0,K^+K^-,K^0{\bar{K}}^0\right)$. $\Pi_r^{ab}$ stands for the diagonal matrix of the polarization operator of the resonance $r$ corresponding to the one loop contribution from the two-particle intermediate states $ab$~\cite{Achasov:2017ncx,Achasov:2018grq},

\indent for $q^2\geq(m_a+m_b)^2$, we have
\begin{align}\label{eq:decaywidthoffamixng4}
\Pi_r^{ab}(q^2)=\frac{g_{rab}^2}{16\pi}\left[\frac{{m_{ab}^{(+)}}{m_{ab}^{(-)}}}{\pi q^2} \ln{\frac{m_b}{m_a}}+\rho_{ab}(q^2)\left(i-\frac{1}{\pi}\ln{\frac{\sqrt{q^2-{m_{ab}^{(-)2}}}+\sqrt{q^2-{m_{ab}^{(+)2}}}}
{\sqrt{q^2-{m_{ab}^{(-)2}}}-\sqrt{q^2-{m_{ab}^{(+)2}}}}}\right)\right];
\end{align}
\indent for $(m_a-m_b)^2<q^2<(m_a+m_b)^2$,
\begin{align}
\Pi_r^{ab}(q^2)=\frac{g_{rab}^2}{16\pi}\left[\frac{{m_{ab}^{(+)}}{m_{ab}^{(-)}}}{\pi q^2} \ln{\frac{m_b}{m_a}}-\rho_{ab}(q^2)\left(1-\frac{2}{\pi} \arctan{\frac{\sqrt{{m_{ab}^{(+)2}}-q^2}}{\sqrt{q^2-{m_{ab}^{(-)2}}}}} \right)\right];
\end{align}
\indent for $q^2\leq(m_a+m_b)^2$,
\begin{align}
\Pi_r^{ab}(q^2)=\frac{g_{rab}^2}{16\pi}\left[\frac{{m_{ab}^{(+)}}{m_{ab}^{(-)}}}{\pi q^2} \ln{\frac{m_b}{m_a}}+\rho_{ab}(q^2)\frac{1}{\pi}\ln{\frac{\sqrt{{m_{ab}^{(+)2}}-q^2}+\sqrt{{m_{ab}^{(-)2}}-q^2}}
{\sqrt{{m_{ab}^{(+)2}}-q^2}-\sqrt{{m_{ab}^{(-)2}}-q^2}}}\right],
\end{align}
where $g_{rab}$ is the coupling of resonance $r$ to final states $ab$, ${m_{ab}^{(\pm)}}=\left|m_a\pm m_b\right|$, and $\rho_{ab}(q^2)$ is
\begin{align}\label{eq:decaywidthoffamixng5}
\rho_{ab}(q^2)=\frac{\sqrt{\left|q^2-{m_{ab}^{(+)2}}\right|}\sqrt{\left|q^2-{m_{ab}^{(-)2}}\right|}}{q^2}.
\end{align}
As for the contribution of $\eta-\pi^0$ mixing, the transition amplitude is
\begin{align}\label{eq:decayamplifpietamixng}
&{\mathcal M}( Y(2175)\rightarrow\phi f_0(980)\rightarrow\phi \pi^0\pi^0\rightarrow\phi\eta\pi^{0})=2{\mathcal M}(Y(2175)\to\phi f_0(980))\cdot \frac{g_{f_0\pi^0\pi^0}}{D_{f_0}(q^2)}\cdot\frac{\lambda_{\pi^0\eta}}{m_{\eta}^2-m_{\pi^0}^2}.
\end{align}
Adding Eq.(\ref{eq:decayamplitudeoffamixng}) and Eq.(\ref{eq:decayamplifpietamixng}) together, we then arrive at
\begin{align}\label{eq:totaldecayamfapietamixng}
&{\mathcal M}( Y(2175)\rightarrow\phi f_0(980)\rightarrow\phi\eta\pi^{0})=\nonumber\\
&~~~~~~={\mathcal M}(Y(2175)\to\phi f_0(980))\cdot\left[\frac{\Pi_{a_0 f_0}(q^2)\cdot g_{a_0\eta\pi^0}}{D_{a_0}(q^2)D_{f_0}(q^2)-\Pi_{a_0 f_0}^2(q^2)}
+ \frac{2g_{f_0\pi^0\pi^0}}{D_{f_0}(q^2)}\cdot\frac{\lambda_{\pi^0\eta}}{m_{\eta}^2-m_{\pi^0}^2}\right],
\end{align}
where $g_{a_0\eta\pi^0}$ and $g_{f_0\pi^0\pi^0}$ are the couplings of $a_0^0(980)$ to $\eta\pi^0$ and $f_0(980)$ to $\pi^0\pi^0$, respectively, which can be extracted from
\begin{align}\label{eq:branchratioforafpieta}
&{\mathcal B}(r\rightarrow a b)=
\frac{g_{rab}^2}{16\pi m_{r}^3\Gamma_{r}} f(m_{r},m_{a},m_{b}).
\end{align}
By combining Eqs.(\ref{eq:ratiofetapipipi2}), (\ref{eq:decayamplitudeofyfphi}), (\ref{eq:totaldecayamfapietamixng}) and (\ref{eq:branchratioforafpieta}), we can obtain the distribution of the $\eta\pi^0$ mass squared spectrum for $J/\psi \rightarrow\eta Y(2175))\rightarrow\eta\phi f_0(980)\rightarrow\eta\phi \eta\pi^{0}$, i.e.
\begin{align}\label{eq:decaywidthoffapietamixng1}
{\mathcal B}(J/\psi & \rightarrow\eta Y(2175))\cdot  \frac{d\Gamma(Y(2175)\rightarrow\phi f_0(980)\rightarrow\phi\eta\pi^{0})}{dq^2}=\nonumber\\
&={\mathcal B}(J/\psi \rightarrow\eta Y(2175)\rightarrow\eta\phi f_0(980)\rightarrow\eta\phi \pi^+\pi^-)    \cdot\varphi_{S}     \cdot\left|\delta_{f_0 a_0^0}+\delta_{\pi^0\eta}\right|^2,
\end{align}
where $\varphi_{S}$ is the phase-space factor of the involved decays
\begin{align}\label{eq:decaywidthoffapietamixng2}
\varphi_{S}=\frac{\Gamma_{Y}}{\pi q^2}\cdot\frac{f(m_{Y},m_{\phi},\sqrt{q^2})}{f(m_{Y},m_{\phi},m_{f_0})} \cdot f(\sqrt{q^2},m_{\eta},m_{\pi^0}),
\end{align}
here $\Gamma_Y$ is the total decay width of $Y(2175)$.
$\delta_{\pi^0\eta}$ and $\delta_{f_0 a_0^0}$ in Eq. (\ref{eq:decaywidthoffapietamixng1}) denote the contributions from the $\pi^0$-$\eta$ mixing and the $f_0(980)$-$a_0^0(980)$ mixing,
 respectively, which are given in the following
\begin{align}\label{eq:decaywidthoffapietamixng3}
\delta_{\pi^0\eta}=-\frac{\sqrt{2}}{D_{f_0}(q^2)}  \sqrt{\frac{{\mathcal B}(\eta^{\prime}\rightarrow \pi^+\pi^-\pi^0)}{{\mathcal B}(\eta^{\prime}\rightarrow\pi^+\pi^-\eta)}       \frac{{\phi_s}(\eta^{\prime}\rightarrow\pi^+\pi^-\eta)}{{\phi_s}(\eta^{\prime}\rightarrow\pi^+\pi^-\pi^0)}} \sqrt{\frac{\Gamma_{f_0} m_{f_0}^3}{f(m_{f_0},m_{\pi^0},m_{\pi^0})}},
\end{align}
here, the minus sign is associated with the $\lambda_{\pi^0\eta}$ vertex corresponding to the $\pi^0\leftrightarrow \eta$ transition~\cite{Kudryavtsev:2001ee,Jones:1979ez,Coon:1981qs}.
\begin{align}\label{eq:decaywidthoffapietamixng4}
\delta_{f_0 a_0^0}=\sqrt{\frac{{\mathcal B}(a_0^0(980)\rightarrow \eta\pi^0)}{{\mathcal B}(f_0(980)\rightarrow \pi^+\pi^-)}}\cdot\sqrt{\frac{\Gamma_{a_0} m_{a_0}^3}{f(m_{a_0},m_{\eta},m_{\pi^0})}} \cdot\frac{\Pi_{a_0f_0}(q^2)}{D_{a_0}(q^2)D_{f_0}(q^2)-\Pi_{a_0 f_0}^2(q^2)},
\end{align}
where $\Gamma_{f_0}$ and $\Gamma_{a_0}$ are the decay widths of $f_0(980)$ and $a_0^0(980)$, respectively.
From Refs.~\cite{Cheng:2005nb} and \cite{Cheng:2013fba}, the branching fractions  ${\mathcal B}( f_0 (980)\rightarrow \pi^+ \pi^-)$ and ${\mathcal B}(a_0^0(980)\rightarrow\eta\pi^{0})$ are obtained as
\begin{align}\label{branchingratioofa980}
{\mathcal B}( f_0 (980)\rightarrow \pi^+ \pi^-)=0.50^{+0.07}_{-0.09},\\
{\mathcal B}(a_0^0(980)\rightarrow\eta\pi^{0})=0.845\pm 0.017.
\end{align}
Using the input parameters listed in Table~\ref{massofpietak} and Table~\ref{inputmassparameter}, we obtain the result for the distribution curve of the $\eta\pi^0$ mass squared spectrum for $J/\psi \rightarrow\eta Y(2175))\rightarrow\eta\phi f_0(980)\rightarrow\eta\phi \eta\pi^{0}$ decay, which is shown in Fig.~\ref{maffsquaredistribution}. In this figure, the narrow peak due to the $f_{0}(980)$-$a_0^0(980)$ mixing can be clearly observed.
\begin{table}[t]
\begin{center}
\caption{\label{inputmassparameter} \small Properties of the resonances, here, $f_0$, $a_0$ and $Y$ denote $f_{0}(980)$, $a_0^0(980)$ and $Y(2175)$, respectively. }
\vspace{0.2cm}
\doublerulesep 0.8pt \tabcolsep 0.18in
\begin{tabular}{cc}
\hline
$m_{f_0}=  (0.99\pm 0.02) GeV$ ~\cite{Tanabashi:2018oca}                         &${\Gamma}_{f_0}=  0.074 GeV$~\cite{Ablikim:2017auj}  \\
\hline
$m_{a_0}=  (0.98\pm 0.02) GeV$  ~\cite{Tanabashi:2018oca}                        &${\Gamma}_{a_0}=  (0.092\pm 0.008) GeV$~\cite{Tanabashi:2018oca}     \\
\hline
$m_{Y}=  (2.188\pm 0.010) GeV$~\cite{Tanabashi:2018oca}               &${\Gamma}_{Y}=  (0.083\pm 0.012) GeV$~\cite{Tanabashi:2018oca} \\
\hline
$m_{\phi}=  1019 MeV$~\cite{Tanabashi:2018oca}                                  &$g_{a_0\eta\pi^0}=2.43 GeV$  ~\cite{Cheng:2005nb,Cheng:2013fba}          \\
\hline
$g_{a_0 K^{+}K^{-}}=(2.76\pm 0.46) GeV$~\cite{Kornicer:2016axs,Ablikim:2018ffp}                 &$g_{a_0 K^{0}{\bar{K}}^{0}}=(2.76\pm 0.46) GeV$~\cite{Kornicer:2016axs,Ablikim:2018ffp}\\
\hline
$g_{f_0\pi^{+}\pi^{-}}=1.39 GeV$~\cite{Cheng:2005nb,Cheng:2013fba}                        &$g_{f_0\pi^{0}\pi^{0}}=0.98 GeV$~\cite{Cheng:2005nb,Cheng:2013fba}           \\
\hline
$g_{f_0 K^{+}K^{-}}=3.17 GeV$~\cite{Achasov:2017ncx}                             &$g_{f_0 K^{0}{\bar{K}}^{0}}=3.17 GeV$~\cite{Achasov:2017ncx}      \\
\hline
\end{tabular}
\end{center}
\end{table}

Furthermore, the branching fraction of the decay $J/\psi \rightarrow\eta Y(2175)\rightarrow\eta\phi f_0(980)\rightarrow\eta\phi \eta\pi^{0}$ is obtained by performing the integration in the effective region $(m_{\eta}+m_{\pi^0})^2 \leq q^2 \leq (m_{Y}-m_{\phi})^2$,  and the result is
\begin{align}\label{branchingratioofthetotal}
{\mathcal B}( J/\psi \rightarrow\eta Y(2175)\rightarrow\eta\phi f_0(980)
\rightarrow\eta\phi \eta\pi^{0})=\left(1.30^{+0.67}_{-0.88}\right)\times 10^{-6},
\end{align}
where we have considered  the errors of the mass and width of $a_0^0(980)$ and $f_0(980)$, the errors from the branching fractions of the decays $f_0 (980)\rightarrow \pi^+ \pi^-$ and $a_0^0(980)\rightarrow\eta\pi^{0}$ as well as uncertainty from the branching fraction of the decay $J/\psi \rightarrow\eta Y(2175))\rightarrow\eta\phi f_0(980)\rightarrow\eta\phi\pi^+\pi^{-}$. The contribution from the $f_{0}(980)$-$a_0^0(980)$ mixing dominates the  predicted branching fraction. In additional, the interference of the amplitudes from the $f_{0}(980)$-$a_0^0(980)$ mixing  and $\pi^0 -\eta$ mixing is destructive, the branching fraction is decreased by about $10\%$ owing to the interference effect.

\section{Prospects for the measurement at BESIII}
\label{sec:Prospects for the measurement at BESIII}
The final states $\pi^0$, $\eta$ and $\phi$ in the cascade decay process $J/\psi \rightarrow\eta Y(2175))\rightarrow\eta\phi f_0(980)\rightarrow\eta\phi\eta\pi^{0}$ are reconstructed through the decays $\pi^0\rightarrow\gamma\gamma$, $\eta\rightarrow\gamma\gamma$ and $\phi\rightarrow K^+ K^-$. By employing the data reported by the Particle Data Group~\cite{Tanabashi:2018oca}, we obtain
\begin{align}\label{eq:finalstratebranchratio}
&{\mathcal B}(\eta\rightarrow\gamma\gamma)\cdot{\mathcal B}(\phi\rightarrow K^+ K^-)\cdot{\mathcal B}(\eta\rightarrow\gamma\gamma)\cdot{\mathcal B}(\pi^0\rightarrow\gamma\gamma)=(7.55\pm0.09)\times 10^{-2}.
\end{align}
Because of the narrow peak near the $K\bar{K}$ thresholds in the $\eta\pi^0$ invariant mass spectrum, the event selection criteria for the $a_0^0 (980)$ candidates has high efficiency. In addition, the final states contain six photons and two charged tracks, the detection efficiency for $J/\psi \rightarrow\eta Y(2175))\rightarrow\eta\phi f_0(980)\rightarrow\eta\phi\eta\pi^{0}$ decay can be as large as $8\%$ after the final selection \cite{Ablikim:2014pfc,Ablikim:2015cob,Ablikim:2016frj,Asner:2008nq}. The BESIII experiment will accumulate huge data sample of $10\times 10^9$ $J/\psi$ decays by the end of 2019~\cite{Asner:2008nq,Li:2016tlt,Bigi:2017eni}. Therefore about 80 events for the decay of $J/\psi \rightarrow\eta Y(2175))\rightarrow\eta\phi f_0(980)\rightarrow\eta\phi\eta\pi^{0}$ are expected in the $J/\psi$ decay sample at the BESIII. Therefore, the isospin breaking decay $J/\psi \rightarrow\eta Y(2175)\rightarrow\eta\phi f_0(980)\rightarrow \eta\phi\eta\pi^{0}$ will be helpful to determine the final value of $\xi_{fa}$ in addition to the process $J/\Psi\to \phi f_0(980)\to \phi a_0^0(980)\to\phi\eta\pi^0$.

\section{Conclusions}
\label{sec:Conclusions}
Basing on the branching fraction of the decay $J/\psi \rightarrow\eta Y(2175))\rightarrow\eta\phi f_0(980)\rightarrow\eta\phi\pi^+\pi^{-}$ and the $f_0(980)$-$a_0^0(980)$ mixing intensity $\xi_{fa}$ measured recently by the BESIII, we study the isospin violation decay $J/\psi \rightarrow\eta Y(2175))\rightarrow\eta\phi f_0(980)\rightarrow\eta\phi\eta\pi^{0}$, which proceeds via the $f_0(980)$-$a_0^0(980)$ mixing and the $\pi^0$-$\eta$ mixing. It is found that the decay can reach a branching fraction of the order of $10^{-6}$, which can be accessed  with $10^{10}$ $J/\psi$ events collected at BESIII by the end of 2019. The contribution from the $f_0(980)$-$a_0^0(980)$ mixing dominates the decay. The interference between the amplitude caused by the $f_0(980)$-$a_0^0(980)$ mixing and the amplitude caused by the $\pi^0$-$\eta$ mixing is destructive, the branching fraction will be decreased by about $10\%$ because of the interference effect between the two mixings. In the distribution of the $\eta\pi^0$ mass square spectrum, we find that the narrow peak due to  $f_0(980)$-$a_0^0(980)$ mixing should be expected, and the effect on the peak from $\pi^0-\eta$ mixing is negligibly small.   This decay will  be complementary to the decay $J/\psi\rightarrow\phi f_0(980)\rightarrow\phi\eta\pi^{0}$, which will be helpful to determine the final solution of the $f_0(980)$-$a_0^0(980)$ mixing intensity and understand the nature of the light scalar mesons.
\section*{Acknowledgements}
This work is supported in part by the National Natural
Science Foundation of China under Contracts Nos.~11335009, 11125525,11675137, 11875054,  the Joint Large-Scale Scientific Facility Funds of the NSFC and CAS
under Contract No.~U1532257, CAS under Contract No.~QYZDJ-SSW-SLH003, and the National Key Basic Research Program of China under Contract
No.~2015CB856700.  X.~Cheng and R.~Wang are supported by Nanhu Scholars Program of XYNU.

\begin{appendix}

\end{appendix}

\end{document}